\documentclass[fleqn,usenatbib]{mnras}

\usepackage{newtxtext,newtxmath}

\usepackage[T1]{fontenc}
\usepackage{ae,aecompl}
\usepackage{adjustbox}

\usepackage{graphicx}	
\usepackage{amsmath}	
\usepackage{amssymb}	

\usepackage{tikz}
\usepackage{color}
\usepackage{pgfplots}
\usepackage{pgf-umlsd}
\usepackage{ifthen}
\usepackage{cprotect}
\usepackage[normalem]{ulem}

\usepackage{lscape}
\usepackage{rotating}
\newcommand{\HI}{{\sc Hi}}

\title[Astro-Cosmo constraints with 21cm CNN]{Constraining the astrophysics and cosmology from 21cm tomography using deep learning with the SKA}

\author[Hassan, Andrianomena \& Doughty]{
Sultan Hassan$^{1,2}$\thanks{E-mail: shassan@nmsu.edu}\thanks{Tombaugh Fellow}, Sambatra Andrianomena$^{3,2}$,
Caitlin Doughty$^{1}$
\\
$^{1}$ Department of Astronomy, New Mexico State University, Las Cruces, NM 88003, USA \\
$^{2}$ University of the Western Cape, Bellville, Cape Town 7535, South Africa \\
$^{3}$ South African Radio Astronomy Observatory (SARAO), Black River Park, Observatory, Cape Town, 7925, South Africa \\
}

\date{Accepted XXX. Received YYY; in original form ZZZ}

\pubyear{2015}

\begin{document}
\label{firstpage}
\pagerange{\pageref{firstpage}--\pageref{lastpage}}
\maketitle

\begin{abstract}

Future Square Kilometre Array (SKA) surveys are expected to generate huge datasets of 21cm maps on cosmological scales from the Epoch of Reionization (EoR). We assess the viability of exploiting machine learning techniques, namely, convolutional neural networks (CNN), to simultaneously estimate the astrophysical and cosmological parameters from 21cm maps from semi-numerical simulations. We further convert the simulated 21cm maps into SKA-like mock maps using the detailed SKA antennae distribution, thermal noise and a recipe for foreground cleaning. We successfully design two CNN architectures (VGGNet-like and ResNet-like) that are both efficiently able to extract simultaneously three astrophysical parameters, namely the photon escape fraction (f$_{\rm esc}$), the ionizing emissivity power dependence on  halo mass ($C_{\rm ion}$) and the ionizing emissivity redshift evolution index ($D_{\rm ion}$), and three cosmological parameters, namely the matter density parameter ($\Omega_{m}$), the dimensionless Hubble constant ($h$), and the  matter fluctuation amplitude ($\sigma_{8}$), from 21cm maps at several redshifts. With the presence of noise from SKA, our designed CNNs are still able to recover these astrophysical and cosmological parameters with great accuracy ($R^{2} > 92\%$), improving to $R^{2} > 99\%$ towards low redshift and low neutral fraction values. Our results show that future 21cm observations can play a key role to break  degeneracy between models and tightly constrain the astrophysical and cosmological parameters, using only few frequency channels.
\end{abstract}




\section{Introduction}
The last global phase transition in the Universe, known as the Epoch of Reionization (EoR), marks the time at which the first stars gradually reionized the Inter-Galactic Medium (IGM) and the Universe transitioned from highly neutral-opaque to a highly ionized-transparent state~\citep[for a review, see e.g.][]{doi:10.1146/annurev.astro.39.1.19}. This epoch represents a crucial period in the Universe's history, particularly with regard to the formation and evolution of early galaxies.

Constraining the astrophysical and cosmological parameters has been the focus for most observational and theoretical studies. Several techniques have been developed to constrain the cosmological parameters (e.g matter density parameter $\Omega_{m}$ and Hubble constant $H_{0}$) such as using the Cosmic Microwave Background (CMB) anisotropies measurements~\citep[e.g.][]{Hinshaw_2013,refId0}, Sunyaev-Zel'dovich cluster surveys~\citep[e.g.][]{PhysRevD.68.083506}, galaxy clusters in optical and X-ray bands~\citep[e.g.][]{10.1046/j.1365-8711.2001.04728.x}, gamma ray burst X-ray afterglow light curves~\citep[e.g.][]{10.1111/j.1365-2966.2010.17197.x}, lensed GW+EM signals~\citep[e.g.][]{Li_2019}, Ly-$\alpha$ forest power spectrum and COBE-DMR~\citep[e.g.][]{Phillips_2001}, large-scale clustering of SDSS luminous red galaxies~\citep[e.g.][]{10.1111/j.1365-2966.2007.11593.x}, and a joint CMB and weak lensing analysis~\citep[e.g.][]{PhysRevLett.90.221303}. On the other hand, several works have attempted to constrain the astrophysical parameters (e.g. the photon escape fraction, $f_{\rm esc}$, and ionizing emissivity evolution, $\dot{N}_{\rm ion}$), using Ly-$\alpha$ forest measurements~\citep[e.g.][]{10.1093/mnras/stt1610}, Lyman continuum (LyC) radiation from local galaxies~\citep[e.g.][]{2013A&A...553A.106L}, and inferred constraints by tuning different theoretical models to other measurements~\citep[e.g.][]{10.1093/mnrasl/slv134, 10.1093/mnras/stu2668}. 

While all these methods show different levels of success to place constraints on various parameters, tighter constraints are expected to come from the EoR through measurements of the 21cm fluctuations on cosmological scales. With its strong dependence on the ionization and density fields, the 21cm signal carries a wealth of information that is important in order to understand early stages of galaxy formation and evolution. In this light, many radio interferometer experiments, such as the Low Frequency Array~\citep[LOFAR;][]{2013A&A...556A...2V}, the Precision Array for Probing the Epoch of Reionization~\citep[PAPER;][]{2010AJ....139.1468P}, the Murchison Wide field Array~\citep[MWA;][]{2013PASA...30...31B}, the Giant Metrewave Radio Telescope~\citep[GMRT;][]{2011MNRAS.413.1174P}, the Hydrogen Epoch of Reionization Array \citep[HERA;][]{2017PASP..129d5001D} and Square Kilometer Array~\citep[SKA;][]{2013ExA....36..235M} are devoted to detecting reionization in the near future. These growing observational efforts require equivalent efforts in both the theoretical and statistical sides, in order to prepare for extracting all possible information and constrain the cosmological and astrophysical parameters from future 21cm surveys. 

Several studies have already shown that combining the 21cm power spectrum with Markov Chain Monte Carlo (MCMC) analysis is a powerful technique to obtain tighter constraints and break degeneracy between models~\citep[e.g.][]{10.1093/mnras/stv571,2016PhRvD..93d3013L,2016MNRAS.463L..56P,2017MNRAS.468..122H,2019MNRAS.484..933P}. Besides the power spectrum, future 21cm surveys, the SKA in particular, are also expected to generate huge imaging datasets for the 21cm fluctuations on large scales that will contain more information than the power spectrum. Going beyond the power spectrum has been the target of many studies ~\citep[e.g.][]{2005MNRAS.358..968B,2008MNRAS.384.1069B,2015MNRAS.454.1416W, 2018MNRAS.476.4007M}, in which more information can be obtained through investigating the non-gaussian nature of the 21cm signal using higher-order statistics such as the bispectrum. To efficiently use the 21cm information stored in the 2-dimensional 21cm maps, Convolutional Neural Networks (CNNs) have been a very successful deep learning tool to recover the astrophysical parameters during reionization~\citep{2019MNRAS.484..282G}, to learn the reionization history~\citep{2019ApJ...880..110L,mangena19}, to emulate reionization simulations~\citep{Chardin:2019}, and to identify reionization sources from different models~\citep{2019MNRAS.483.2524H}. However, the astrophysical parameter recovery by~\citet{2019MNRAS.484..282G} ignores the instrumental effects as an initial proof-of-concept study. Accounting for these effects such as the angular resolution, foreground cleaning, and thermal noise, are all crucial in order to add realism to the simulated 21cm images as we prepare for the 21cm era.

In this work, we take a step further to design two different CNNs to simultaneously estimate  several parameters from 21cm maps at several redshifts and different stages through reionization. We here that assume all observations at different redshifts are performed independently. We simply take maps from different redshifts and apply the instrumental noise directly on each map assuming a single frequency channel of a size $\sim$ 50 KHz (i.e. simulation resolution). We finally combine the maps from different redshifts to create our training datasets. We note that learning from light-cones is beyond the scope of the current work. Our aim is to provide a network that is able to predict parameters without requiring the redshift nor neutral fraction as inputs, which is a more flexible design. Three astrophysical parameters are evaluated: the photon escape fraction (f$_{\rm esc}$), the ionizing emissivity power dependence on halo mass ($C_{\rm ion}$) and the redshift evolution index ($D_{\rm ion}$). Additionally, we estimate three cosmological parameters: the matter density parameter ($\Omega_{m}$), the dimensionless Hubble constant ($h$), and the matter fluctuation amplitude ($\sigma_{8}$). To assess the ability of future 21cm tomography to constrain these parameters, we follow the recipe presented in~\citet{2019MNRAS.483.2524H} to add a physically motivated and realistic 21cm noise to large scale 21cm maps that are produced using our semi-numerical model, {\sc SimFast21} \citep{santos2008cosmic, santos2010fast}. This paper is organized as follows: we first describe our suite of simulations of the 21cm signal and noise in \S\ref{simulation-data}. We then present the two network designs in \S\ref{networks} and the training dataset in \S\ref{sec:train_set}. We present the main results in \S\ref{results}, and draw our concluding remarks in \S\ref{conclusion}.

\section{simulations}\label{simulation-data}

\subsection{Semi-Numerical Model, {\sc SimFast21}}

We use the Instantaneous model of our semi-numerical simulations {\sc SimFast21}, that has been developed in \cite{2016MNRAS.457.1550H}, to improve over previous implementations of the ionizing source and sink populations in these semi-numerical simulations. In addition, it has been recently shown that this model is in a  relatively good agreement with predictions from our radiative transfer simulation~\citep[{\sc ARTIST};][]{2019arXiv190103340M},  particularly in terms of the morphology and power spectrum of the ionization and 21cm fields. However, the reionization history can be quite different for the same photon escape fraction value. This arises from violation of photon conservation which is an intrinsic problem in the use of excursion set-formalism in semi-numerical simulations~\citep{2007ApJ...654...12Z,2016MNRAS.460.1801P,2017MNRAS.468..122H}. As indicated by {\sc ARTIST}, as a temporary solution all our photon escape fraction predictions can be adjusted by a factor of 20\% to account for the photon conservation problem. We here briefly describe the simulation ingredients, and defer to \cite{santos2010fast} for the full details of the simulation algorithm, and to \cite{2016MNRAS.457.1550H} for the Instantaneous model development.

The dark matter density is generated in the linear regime from a Gaussian distribution using a Monte-Carlo approach. Evolving the density field to non-linear regime is performed through the \cite{1970A&A.....5...84Z} approximation. Halos are then generated using the excursion-set formalism (ESF). Ionized regions are identified using a similar form of the ESF that is based on a direct comparison between the instantaneous rates of ionization $R_{\rm ion}$ and recombination $R_{\rm rec}$ in spherical regions of decreasing sizes as specified by the ESF. Regions are flagged as ionized if:
\begin{equation}
\centering
f_{\rm esc}\,  R_{\rm ion} \geq R_{\rm rec}\, ,
\end{equation}
where $f_{\rm esc}$ is the escape fraction. The $R_{\rm rec}$ is obtained from a radiative transfer simulation \citep{2015MNRAS.447.2526F}, in order to account for the clumping effects below our cell size. The $R_{\rm rec}$ is parameterized as a function of overdensity $\Delta$ and redshfit $z$ as follows:
\begin{equation}
\frac{R_{\rm rec}}{V} =  9.85 \times 10^{-24} (1+ z)^{5.1}  \left[\frac{\left( \Delta/1.76 \right)^{0.82}}{1+ \left( \Delta/1.76 \right)^{0.82} } \right]^{4} \, , 
\end{equation}
where  V refers to the cell volume. The $R_{\rm ion}$ parameterization is derived from a combination of the radiative transfer simulation \citep{2015MNRAS.447.2526F}, and a larger hydrodynamic simulation \citep{2013MNRAS.434.2645D} that both have been shown to reproduce wide range of observations, including low-$z$ observations. The $R_{\rm ion}$ is parameterized as a function of halo mass $M_{\rm h}$ and redshift $z$ as follows:
\begin{equation}\label{eq:nion}
\frac{R_{\rm ion}}{M_{h}} =  1.1\times 10^{40} \,(1 + z)^{D_{\rm ion}} \left( \frac{M_{h}}{9.51\times 10^{7}}\right)^{C_{\rm ion}}\exp\left(  \frac{-9.51\times 10^{7}}{M_{h}}\right)^{3.0} \, ,
\end{equation}
where the best fit values of the ionizing emissivity dependence on halo mass $C_{\rm ion}$ and redshift $D_{\rm ion}$ were found to be $C_{\rm ion} = 0.41$ and $D_{\rm ion} = 2.28$, respectively. Later, we will change these parameters to generate the training and testing datasets. Note that Equation~\eqref{eq:nion} shows that $R_{\rm ion}$ scales as M$_{\rm h}^{1.41}$,  which is consistent with the SFR$-$M$_{\rm h}$ relation previously found by~\citet{2011ApJ...743..169F}.  We defer to \citet{2016MNRAS.457.1550H} for the full details on the derivation of the $R_{\rm ion}$ and $R_{\rm rec}$ fitting functions and their effects on several reionization key observables.
\newline
\subsection{21cm Instrument Simulation}
We here describe the method used to account for various instrumental effects following the recipe developed in~\citet{2019MNRAS.483.2524H}. We briefly review this method below and refer the interested readers to~\citet{2019MNRAS.483.2524H} for detailed information and complete steps of how we convert an 21cm simulated map into a \textit{mock map} according to the assumed array design. In this work, we restrict our analysis to the SKA proposed design and leave a more detailed comparison between different arrays, such as HERA and LOFAR, for future works. The instrumental noise is applied separately on each redshift assuming a single frequency channel corresponding to the map size ($\sim$ 50 kHz). We leave to future works learning from the light-cones by considering many frequency channels over a large bandwidth in the analysis.

\begin{table}\huge
 \scalebox{0.5}{\begin{tabular}{ l  c   }\hline
    Array design & 866 compact core \\ \hline
    Station diameter, $D$ [m] & 35 \\ \hline
    Station area, $A\,\, [m^{2}]$ &  962.11 $\left( \frac{110}{\nu [\rm MHz]} \right)^{2}$ \\ \hline
    System temperature [K] ($T_{\rm sys}=T_{\rm sky} + T_{\rm rcvr}$) &  1.1 $T_{\rm sky}$ + 40  \\ \hline
    Total observation time t$_{\rm int}$ [h] & 1000  \\ \hline
    Frequency resolution $\Delta \nu$ [kHz]  & 48 \\ \hline
    Redshift &  10, 9, 8 ,7  \\ \hline
    Frequency [MHz] &  129 , 142, 158, 178              \\ \hline
    FWHM [arcmin] &  1.37, 1.24, 1.12, 0.99   \\ \hline
    Beam angle $\theta$ [rad]  &  0.066, 0.06, 0.054, 0.048   \\ \hline
    Default wedge slope $m$, Equation.~\eqref{eq:wedge_eq} & 0.27, 0.23, 0.19, 0.15\\ \hline
\end{tabular}}
\caption{Summary of our assumed SKA array design.}\label{array_tab}
\end{table}

\begin{figure*} 
    \centering 
    \includegraphics[scale=0.6]{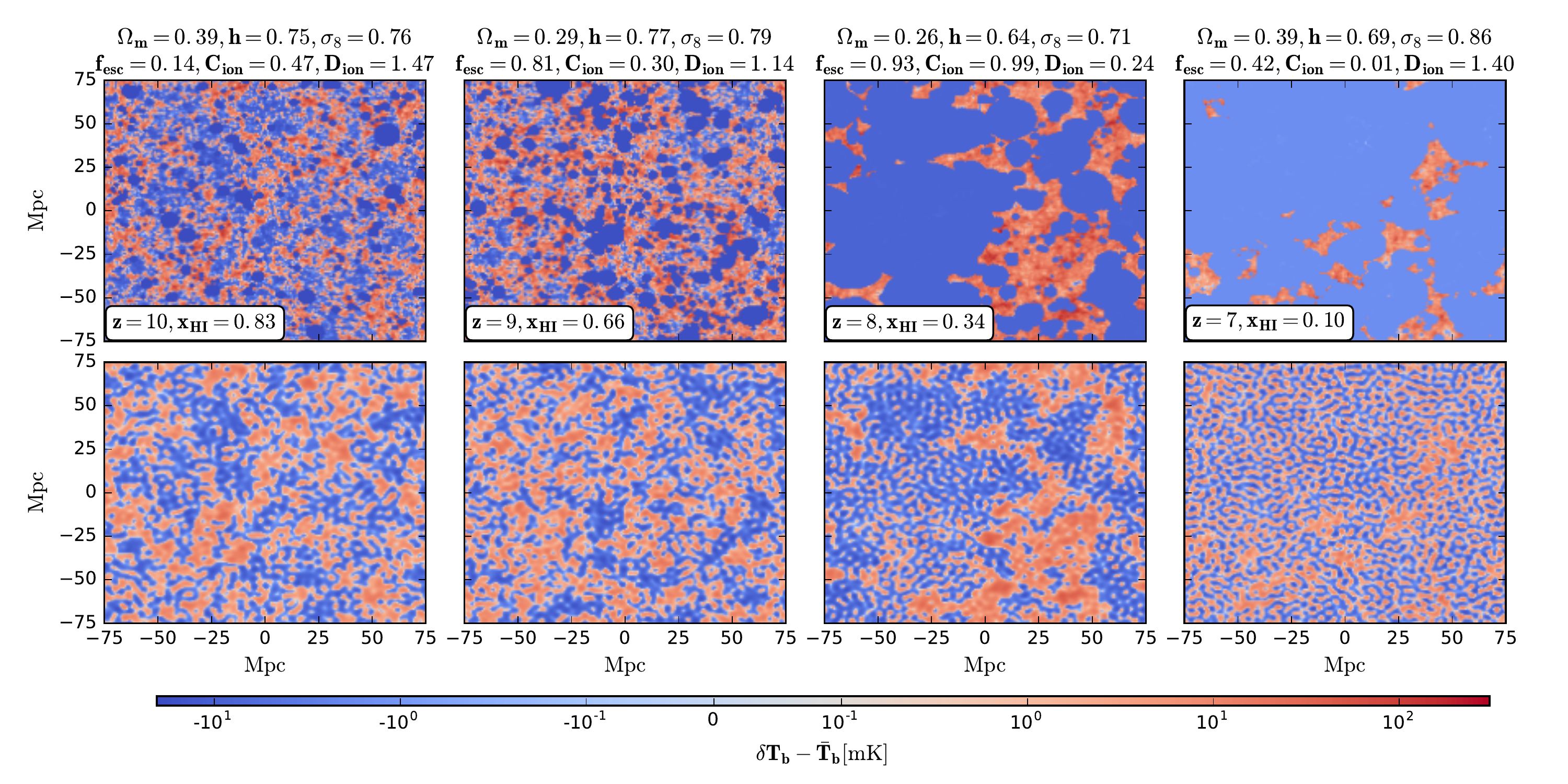}
    \caption{Examples of four randomly selected 21cm maps (top), from our training dataset, with their corresponding mock version (bottom), using our assumed SKA design. Red and blue color represent neutral and ionized regions, respectively. Subtitles show the astrophysical and cosmological parameters used to generated each map. These parameters are: 
   the photon escape fraction (f$_{\rm esc}$), ionizing emissivity power dependence on  halo mass ($C_{\rm ion}$) and ionizing emissivity redshift evolution index ($D_{\rm ion}$),  matter density parameter ($\Omega_{m}$),  dimensionless Hubble constant ($h$), and matter fluctuation amplitude ($\sigma_{8}$). Coloured version is available online.}
    \label{fig:samples}
\end{figure*}
The 21cm Instrument simulation pipeline consists of three parts:
\begin{itemize}
\item {\bf Foreground cleaning:} Foreground contaminated modes of the signal lie inside the foreground wedge in the $k_{\perp}-k_{\parallel}$ plane. The foreground wedge slope (m) is given by:
\begin{equation}\label{eq:wedge_eq} 
m = \frac{ D \, H_{0}\, E(z)\, \sin \theta}{c(1+z)},
\end{equation}
where $H_{0}$ is the Hubble parameter, $c$ is the speed of light, $E(z)\equiv\sqrt{\Omega_{m}(1+z)^{3} +\Omega_\Lambda}$, and $\theta$ is the beam angle. To clean foregrounds, we simply zero out all modes within the wedge, satisfying k$_{\parallel} <$ m k$_{\perp}$. For the same experiment, the slope increases with redshift, which means more modes are removed at higher redshifts. We quote exact wedge slope values for the SKA at our redshifts of interest in Table~\ref{array_tab}. This the first step of the noise pipeline to clean foregrounds from the 3-dimensional co-eval cubes.

\item {\bf Angular resolution:} we account for the angular resolution of a given array by exploiting its detailed baseline distribution, via the $uv$-coverage, which is a measure of the baseline intensity observing the signal modes in directions perpendicular to the sightline. The $uv$-coverage is computed using the {\sc 21cmSense} package\footnote{https://github.com/jpober/21cmSense} from our assumed SKA antennae distribution. We then Fourier transform the simulated 21cm map and set the signal to zero at k$_{\perp}$ modes whose $uv$-coverage is zero\footnote{Modes with zero $uv$-coverage  lie outside the angular resolution of the experiment.}. We additionally smooth down the simulated maps using a Gaussian filter whose full width half maximum (FWHM) is given by: FWHM = $\lambda_{21cm}$ (1+z)/B, whereas the maximum baseline length B=5,834 m for our assumed SKA design, and $\lambda_{21cm}$ is the rest frame wavelength of the 21cm signal. This sets the minimum angular resolution for our assumed SKA design. For instance, our simulated maps initially have an angular resolution of $\sim$ 0.3$^\prime$ at z=7, that are smoothed to have a lower angular resolution of $\sim$ 1$^\prime$ according to the FWHM at this redshift. Exact angular resolution values as a function of redshift are quoted in Table~\ref{array_tab}. The angular resolution recipe is applied on maps extracted from the 3-dimensional foreground filtered boxes from the previous step.

\item {\bf Thermal noise:} the thermal noise is uncorrelated between measurements, and can be drawn from a Gaussian distribution of unit mean and standard deviation \citep{2004ApJ...608..622Z} given by:
\begin{equation}
\sqrt{\langle|N|^{2}\rangle} {\rm [Jy]}  =\frac{ 2 \,k_{\rm B}\, T_{\rm sys} }{ A \, \sqrt{\Delta\nu \,t_{\rm int}}  } \, ,
\end{equation}
where t$_{\rm int}$ here is the integration time to observe a single visibility at a frequency resolution $\Delta \nu$, and $k_{\rm B}$ is the Boltzmann constant. The total system temperature T$_{sys}$ and other parameters are summarized in Table~\ref{array_tab}. Having generated the thermal noise in 2D grid using the above equation in Fourier space, we further suppress the noise by the amount of the $uv$-coverage $N_{uv}$ by a factor of $\sim 1/\sqrt{ N_{uv} }$. We finally inverse Fourier transform the noise map and add it to the angular resolution - foreground filtered signal map to form our mock 21cm map. 
\end{itemize}
Using this pipeline with parameters listed in Table~\ref{array_tab}, the rms brightness temperature (noise level) is about $\sim$ 3 mK at z=8, consistent with previous estimates~\citep[e.g. see][]{2006PhR...433..181F,2017MNRAS.471.1936K,2018MNRAS.479.5596G}.
This pipeline is used to add realism to our simulated training and testing dataset, in order to assess the ability of future SKA 21cm surveys to constrain the astrophysical and cosmological constraints. In Figure~\ref{fig:samples}, we show an example of four randomly selected 21cm maps (top) with their mock versions (bottom) from our training datatset for different set of astrophysical and cosmological parameters as quoted in the subtitles. These maps are generated from different simulations realizations of a box size of L=150 Mpc, number of cells N=200, resulting in a resolution of 0.75 Mpc.  We find that most of the large and small scale ionized bubbles are still present after adding the instrumental effects. This is due to the high angular resolution of our assumed SKA design as well as the high $uv$-coverage that extends down to a very small scales ($\sim$ 3.5 h/Mpc) during these epochs.

However, fully ionized ($x_{\rm HI} < 0.01$) and fully neutral maps ($x_{\rm HI} > 0.99$), as described later in \S\ref{sec:train_set}, are already excluded from the training sample, since they are identical for different set of parameters. Distinguishing identical maps is challenging for neural networks, where more information, such as redshift evolution, is required to assist parameter recovery at these extreme limits.   When the Universe is highly ionized (e.g. $x_{\rm HI} \sim 0.1-0.2$, see column 4 in Figure~\ref{fig:samples}), the noise dominates but nevertheless the residual neutral patches can still be seen and recognized. These residual patches are usually different for different set of parameters, which might help the network to distinguish between maps and parameters.  On the other hand, in the beginning of reionization, the ionized regions are very small due to the small number of sources and ionizing photons. The noise then contaminates and fills these small ionized regions (e.g. $x_{\rm HI} \sim 0.8-0.9$, see column 1 in Figure~\ref{fig:samples}), and hence maps might look similar to those from a fully neutral Universe. This makes recognizing the prominent signal features more challenging, and many of the reionization realizations for a highly neutral Universe become approximately indistinguishable.
This might impact the parameter recovery from a highly neutral IGM, which basically exists at high redshifts where the noise is stronger.

\section{Network architecture}\label{networks}

\begin{table}
 \caption{The architecture of \textbf{Network I} for this study.}
 \label{setups}
 \begin{tabular}{lll}
  \hline
   & Layer & Output shape \\
  \hline
  \hline
  1  & Input & (1, 200, 200)\\[2pt]
  2  & 3$\times$3 Convolutional Layer & (32, 200, 200)\\[2pt]
  3  & 3$\times$3  Convolutional Layer & (32, 200, 200)\\[2pt]
  4  & Batch Normalization & --\\[2pt]
  5  & ReLU Activation & --\\[2pt]
  6  & 2$\times$2  Max Pooling & (32, 100, 100)\\[2pt]
  7  & 3$\times$3  Convolutional Layer & (64, 100, 100)\\[2pt]
  8  & 3$\times$3  Convolutional Layer & (64, 100, 100)\\[2pt]
  9  & Batch Normalization & --\\[2pt]
  10 & ReLU Activation & --\\[2pt]
  11 & 2$\times$2 Max Pooling & (64, 50, 50)\\[2pt]
  12 & 3$\times$3  Convolutional Layer & (128, 50, 50)\\[2pt]
  13 & 3$\times$3  Convolutional Layer & (128, 50, 50)\\[2pt]
  14 & Batch Normalization & --\\[2pt]
  15 & ReLU Activation & --\\[2pt]
  16 & 2$\times$2 Max Pooling & (128, 25, 25)\\[2pt]
  17 & 3$\times$3  Convolutional Layer & (256, 25, 25)\\[2pt]
  18 & 3$\times$3  Convolutional Layer & (256, 25, 25)\\[2pt]
  19 & Batch Normalization & --\\[2pt]
  20 & ReLU Activation & --\\[2pt]
  21 & $\rm{Fully\ Connected\ Layer}$ &  (1024)\\[2pt]
  22 & Batch Normalization & --\\[2pt]
  23 & ReLU Activation & --\\[2pt]
  24 & $\rm{Fully\ Connected\ Layer}$ &  (1024)\\[2pt]
  25 & Batch Normalization & --\\[2pt]
  26 & ReLU Activation & --\\[2pt]
  27 & $\rm{Fully\ Connected\ Layer}$ &  (1024)\\[2pt]
  28 & Batch Normalization & --\\[2pt]
  29 & ReLU Activation & --\\[2pt]
  30 & $\rm{Fully\ Connected\ Layer}$ &  (6)\\[2pt]
  \hline
 \end{tabular}
 \label{network1}
\end{table}

\begin{table}
 \caption{The architecture of \textbf{Network II} for this study.}
 \label{setups}
 \begin{tabular}{lll}
  \hline
   & Layer & Output shape \\
  \hline
  \hline
  1  & Input & (1, 200, 200)\\[2pt]
  2  & Convolutional Layer & (16, 200, 200)\\[2pt]
  3 & Batch Normalization & -- \\[2pt]
  4  & ReLU Activation & -- \\[2pt]
  5 & Residual Layer (3 Residual Blocks) &  (16, 100, 100)\\[2pt]
  6 & Residual Layer (6 Residual Blocks) &  (32, 50, 50)\\[2pt]
  7 & Residual Layer (6 Residual Blocks) &  (64, 25, 25)\\[2pt]
  8 & Residual Layer (3 Residual Blocks) &  (128, 13, 13)\\[2pt]
  9 & Inception Module &  (240, 13, 13)\\[2pt]
  10 & Max Pooling &  (240, 7, 7)\\[2pt]
  11 & Inception Module &  (240, 7, 7)\\[2pt]
  12 & Inception Module &  (256, 7, 7)\\[2pt]
  13 & Inception Module &  (288, 7, 7)\\[2pt]
  14 & Max Pooling &  (288, 4, 4)\\[2pt]
  15 & $\rm{Fully\ Connected\ Layer}$ &  (1024)\\[2pt]
  16 & Batch Normalization & -- \\[2pt]
  17 & ReLU Activation &  -- \\[2pt]
  18 & $\rm{Fully\ Connected\ Layer}$ &  (1024)\\[2pt]
  19 & Batch Normalization & -- \\[2pt]
  20 & ReLU Activation &  -- \\[2pt]
  21 & $\rm{Fully\ Connected\ Layer}$ &  (1024)\\[2pt]
  22 & Batch Normalization & -- \\[2pt]
  23 & ReLU Activation &  -- \\[2pt]
  24 & $\rm{Fully\ Connected\ Layer}$ &  (6)\\[2pt]
  \hline
 \end{tabular}
 \label{network2}
\end{table}

We consider two types of network in this work. It is worth reiterating that our main objective is to able to infer both astrophysical $ \{ f_{\rm esc}, C_{\rm ion}, D_{\rm ion}\}$ and cosmological $\{ \Omega_{\rm m} , \textit{h} , \sigma_{8} \}$ parameters simultaneously from their corresponding 21cm maps. To this end, our main focus is to simply explore different network designs with different layout in width and depth as an attempt to achieve our goal.

\begin{figure}
    \centering
    \includegraphics[scale=0.6]{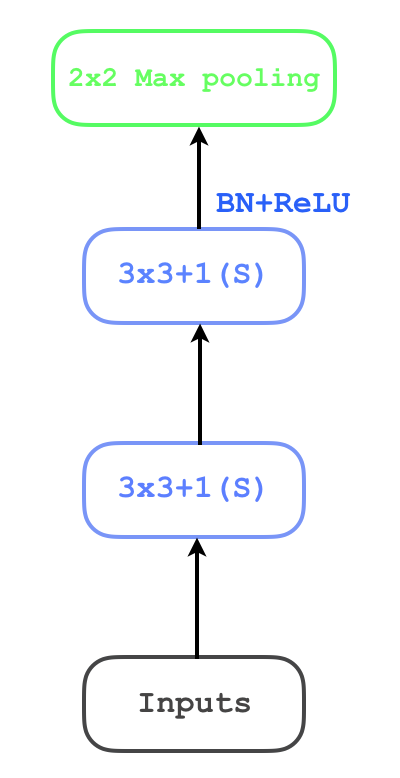}
    \caption{One stage in \textbf{network I}. Chaining two convolutional layers with same number of feature maps followed by a batch normalisation and ReLU function before a max pooling. Coloured version is available online.}
    \label{fig:stage}
\end{figure}

The first architecture (\textbf{network I}) considered for our investigation is given in Table.~\ref{network1}. It is slightly similar to \verb|VGG-Net| \citep{simonyan2014very} in terms of chaining convolutional layers before downsampling, however the key difference here is that each stage\footnote{which we refer to as mapping the input $\textbf{x}$ without reducing the dimensions ($\rm{weight} \times \rm{height}$).}, we have two convolutional layers with same number of feature maps in a row followed by batch normalisation and ReLU activation (\verb|Conv+Conv+BN+ReLU|) as shown in Figure~\ref{fig:stage}. We note that the representation \verb|N x N + M(S)| denotes the kernel size (\verb|N x N|) and the stride (\verb|M|) of a convolutional layer. In total, we have four stages, each with a \verb|Conv+Conv+BN+ReLU| layer followed by a \verb|Max Pooling| with stride = 2 to reduce the dimensions of the inputs\footnote{In other words, the ouputs from the previous stage.}, and four dense layers each with 1024 units with the exception of the output layer, which has only 6 units corresponding to the number of inferred parameters. This network design is also similar to the previous design used in our reionization models classifier~\citep{2019MNRAS.483.2524H}, except that the convolutional layers used here are wider and no dropout seems to be needed. This is consistent with the disharmony observed between batch normalization and dropout~\citep{2018arXiv180105134L}. Similar to our previous works in the classifier, we initialize the network weights using 
a generalized form of Xavier initializer \citep{glorot2010understanding} that
is also called the Variance Scaling initializer, in which the random numbers are drawn from a zero mean Gaussian distribution whose variance is equal to the inverse of the average of the number of input and output neurons. This initializer ensures that the variance of the input data is preserved as it propagates through the network layers.

\begin{figure}
    \centering
    \includegraphics[scale=0.45]{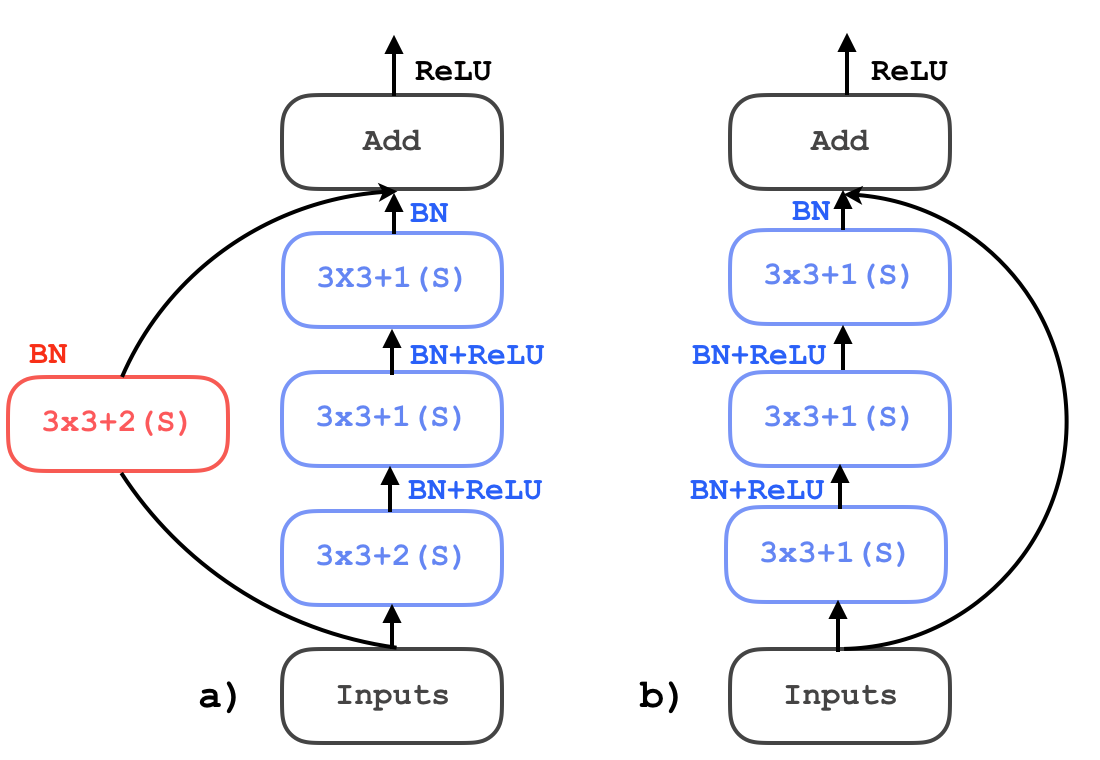}
    \cprotect\caption{Residual block in \textbf{network II}. \textit{Left panel}: the downsampling only occurs at the first convolutional layer (blue \verb|3x3+2(S)|) but the dimension  is kept the same at the second convolutional layer (blue \verb|3x3+1(S)|).  To match the dimensions of the output from the chain of convolutional layers (blue ones) the input is fed to a convolutional layer with strides = 2 (red \verb|3x3+2(S)|). \textit{Right panel}: when there is no downsampling, the input is simply added to the output from the chain of convolutional layers (blue ones).  Coloured version is available online.}
    \label{fig:resnet}
\end{figure}

Our second architecture, which we simply name \textbf{network II}, is based on a combination of residual layers \citep{he2016deep} and inception modules \citep{szegedy2015going} as shown in Table.~\ref{network2}. The inputs, as described in \S\ref{simulation-data}, are first fed into a convolutional layer followed by a batch normalization \citep{ioffe2015batch} before a ReLU activation (\verb|Conv+BN+ReLU|). This is then followed by four residual layers, each composed of three, six, six and three residual blocks respectively.  It was shown in \cite{he2016deep} that the resulting error (both training and testing) of deeper architecture tends to be larger than that of shallower architecture. Therefore they proposed a residual layer which allowed them to increase the depth of the model in order to gain better performance. In contrast with {\bf network I}, instead of using simple convolutional layers we stack residual blocks, which are achieved with the schematic shown in Figure.~\ref{fig:resnet} (right panel) where the residual learning is constructed using a \verb|Conv+BN+ReLU+Conv+BN+ReLU+Conv+BN| layer.
Depending on whether there is downsampling (Fig.~\ref{fig:resnet} left panel) through the chain of convolutional layers in a residual block, the input needs to be downsampled using a \verb|Conv+BN| layer to match the dimension of the output of the chain of convolutional layers.
 
\begin{figure}
    \includegraphics[scale=0.4]{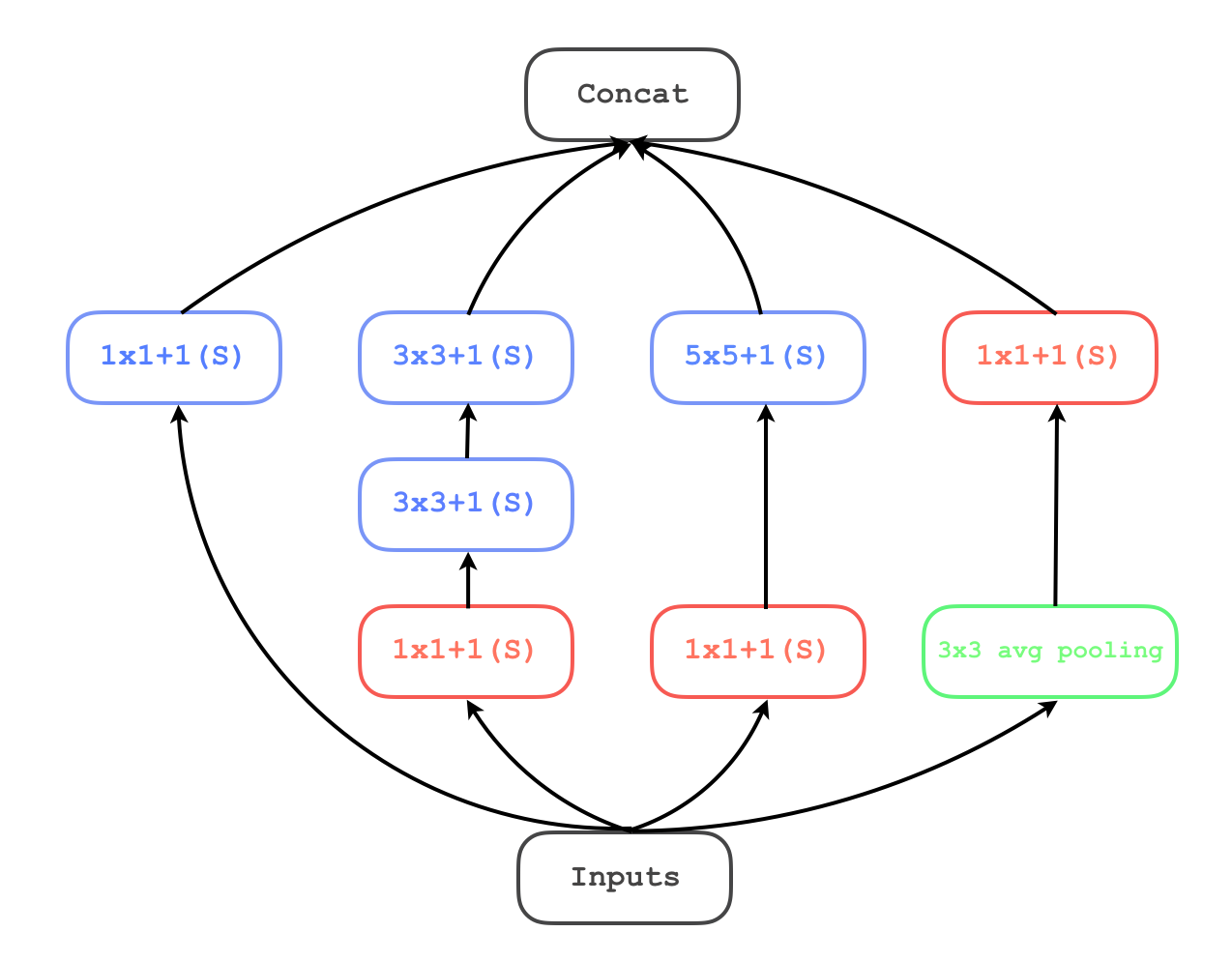}
    \cprotect\caption{Inception module considered in this study. The red convolutional layers (\verb|1x1+1(S)|) are used for dimensionality reduction.  Coloured version is available online.}
    \label{fig:inception}
\end{figure}
In each residual layer, the downsampling occurs at the first residual block. There are variants of deep residual networks, but in essence what we consider here is such that the network performance is optimized for our specific task. 

As proposed by \citet{szegedy2015going}, in order to improve the recognition of more complex features at the higher levels of the network, we make use of four inception modules after the residual layers. The prescription suggested in \citet{szegedy2015going} is to deal with the computational complexity related to the depth of the network, \textit{i.e.} increasing the size of the network while maintaining the computational cost. The inception module used in this network design is shown in Figure.~\ref{fig:inception}. The idea behind convolving the inputs with a $1\times1$ filter before the convolutional layers with $3\times3$ and $5\times5$ kernels is to reduce the number of feature maps from the inputs as computations are more expensive with larger kernel size. The features at different scales -- captured by different kernel sizes $1\times1$, $3\times3$ and $5\times5$ -- can be learned \textit{simultaneously}~\citep{szegedy2015going}. It is worth noting that we opt for He initialization \citep{he2016deep} for all layers in \textbf{network II}. 

For training, as usual for any machine learning tasks, one needs to fine-tune the hyperparameters in order for the algorithm to generalize well and hence achieve the best possible performance, where the distance between the ground truth and network predictions is minimum. To that end,  as shown in Table \ref{optimisers} in Appendix~\ref{sec:app2}, we use two completely different approaches in terms of optimisation for the two architectures. Although the two networks produce similar results, as will be presented in \S\ref{results}, \textbf{network I} converges faster. This can be explained by the capacity of \textbf{network I} with its number of trainable parameters of about 167 millions which translates to $\sim 2.05\times10^9$ floating point operation per second (flops) at inference time, whereas \textbf{network II} has $\sim 10$ millions of trainable parameters corresponding to $\sim 1.39\times10^9$ flops at inference time.

For reproducibility, we have used {\sc TensorFlow} package \citep{tensorflow2015-whitepaper} to develop {\bf network I} which has been trained for 50,000 training steps (about 100 training steps per epoch) which spend $\sim$ 15 hours on a single GPU. Each training step with a batch size of 128 images takes about $\sim$ 1 second.  The network converges from the first few epochs but reaches minimum ({\sc RMSE$\sim$ 0.001}) at epoch = 40 (see Figure~\ref{fig:progress} in Appendix~\ref{sec:app1}). This indicates that same results can be obtained in about 6 hours with a single GPU. For {\bf network II}, we have used {\sc PyTorch} \citep{NEURIPS2019_9015} resorting to three GPUs to speed up the convergence. Each epoch, in which each GPU processes in parallel a batch of 128 images at a time, takes about 2 minutes which is translated to 40 hours for 1200 epochs.

\section{Training dataset}\label{sec:train_set}
We generate the training dataset from a large simulation box of a size L=150 Mpc with N=200$^{3}$ cells. We run 1,000 different reionization simulations realizations with 1,000 different random seeds for the initial density field fluctuations. The prior range assumed to our parameters of interest is as follows:
\begin{itemize}
    \item {\bf Cosmology}: 
    \begin{itemize}
        \item  Matter density parameter: $0.2 \leq \Omega_{m} \leq 0.4$.
        \item  Hubble constant: $0.6 \leq {\it h} \leq 0.8$.
        \item  Matter fluctuation amplitude:  $0.7 \leq \sigma_{8} \leq 0.9$.
    \end{itemize}
    \item {\bf Astrophysics}:
    \begin{itemize}
        \item Photon escape fraction: $0.01 \leq f_{\rm esc} \leq 1$.
        \item R$_{\rm ion}$-M$_{\rm h}$ power dependence:  $0 \leq C_{\rm ion} \leq 1$.
        \item R$_{\rm ion}$ redshift evolution index: $0 \leq D_{\rm ion} \leq 2$.
    \end{itemize}
\end{itemize}
The ranges considered for the astrophysical parameters are motivated from our previous MCMC estimates to reproduce various reionization observables~\citep{2017MNRAS.468..122H}, and those of the cosmology are inspired by the recent parameters estimates from the Planck Collaboration 2018 \citep{aghanim2018planck}. From these priors, we select 1,000 values for each parameter using Latin Hypercube Sampling (LHS) in order to efficiently explore our 6-dimensional parameter space and ensuring that the simulation doesn't run twice using the same set of parameters. From each simulation run, we store the 21cm brightness temperature at several redshifts $z$ = 10, 9, 8, 7 in order to have a sufficiently large number of maps to ensure training.  Balancing the training data set is important to ensure equal learning at all redshifts and all neutral fraction values. This can be achieved by flattening the distribution according to the neutral fraction at each redshift. Flattening the distributions has previously been used in learning  cluster masses~\citep{2015ApJ...803...50N} to reduce the bias towards low mass clusters. To flatten our distribution, we take the following steps: First, we ignore highly ionized ($x_{\rm HI} < 0.01$) and highly neutral ($x_{\rm HI} > 0.99$) 21cm boxes. Second, at each redshift, we bin the boxes according to their neutral fraction. Since the neutral fraction changes strongly with different parameters at different redshifts, the number of boxes in each neutral fraction bin is also different. One has to choose a fixed number to select boxes from bins to flatten the distribution. We here choose 20 boxes. These 20 boxes are randomly selected from each bin. If the neutral fraction bin has less than 20 boxes, then we consider all boxes in this specific bin. If all neutral fraction bins have 20 boxes at the 4 different redshifts, then the total number of all boxes is 800. However, few bins have less than 20 boxes which reduces the total number of boxes to 763. Each selected box has 200 different maps along each of $x,y,z$ directions. This means each redshift has $200\times 3\times 1000 = 600000$ possible different 21cm maps. However, close maps in the same box would contain similar features, and from our own experience~\citep[e.g.][]{2019MNRAS.483.2524H}, we have found that $\sim$ 2 Mpc separation between maps is sufficient to obtain distinct maps. To be more conservative, we consider $\sim$ 4 Mpc separation between maps to only select 40 slices along two directions (e.g. x, y) for training, and take 10 slices on the third direction (e.g. z) for testing and validation each. This results in 763 $\times$ ( 40$\times$2 + 10 + 10) = 76,300 total number of images, in which 80\% is used for training, 10\% for validation and the remaining 10\% for testing. In Figure~\ref{fig:xHI_distribution}, we show the histogram of the training data set as a function of neutral fraction at each redshift. The distribution is approximately flat by construction and includes all possible neutral fraction values at each redshift. We here ignore all current reionization constraints to allow specific neutral fraction values at each redshift such as allowing only high neutral fractions at high redshifts and vice versa. This is an important initial test when constraining parameters, which is to verify the method viability to recover these parameters without imposing any priors and constraints.  It is worthwhile to mention that the time-dependence between maps is included through the following:

\begin{figure}
    \centering
    \includegraphics[scale=0.5]{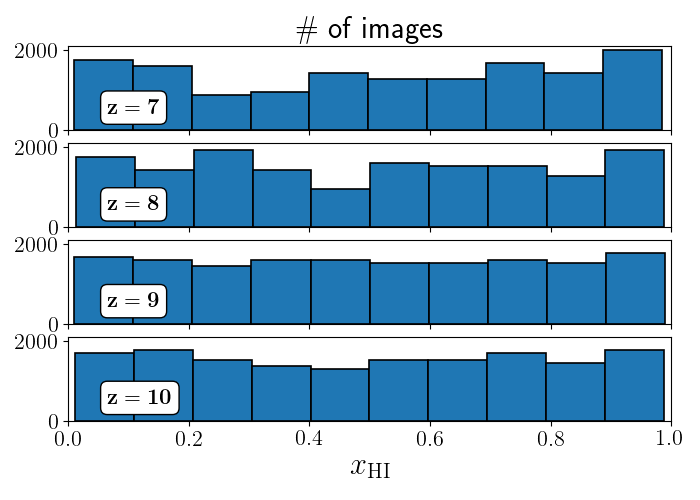}
    \caption{The distribution of training sample at z=7,8,9,10 (top to bottom) as a function of neutral fraction. We intentionally ignore all current reionization history constraints to check the CNNs' viability  to recover parameters without imposing any priors. Coloured version is available online.}
    \label{fig:xHI_distribution}
\end{figure}

\begin{itemize}
    \item Each set of the six parameters corresponds to four boxes at redshifts z=10,9,8,7. This shows that the network sees the same six parameters for four different maps from four different redshifts.
    \item  The redshift information is encoded in each box through the density field contribution on small scales. This shows that the network sees four different levels of density field contribution in the neutral regions in all maps. 
\end{itemize}
However, an explicit inclusion of this effect is through creating light-cones for each set of the six parameters to account for ionized bubbles growth along the sightline (i.e. the k$_{\parallel}$ modes), redshift-space distortion and angular scales. However, we here assume that all observations at different redshifts are performed independently and apply the noise using a single frequency channel (resolution) corresponding to the map size. The light-cone is more relevant when the full bandwidth is considered. This study sets the baseline for a more detailed analysis to compare the results from 2-dimensional maps (single frequency channel) versus  3-dimensional light-cones (full bandwidth). Indeed, it is expected that more information exists in the reionization window~\citep[e.g.][]{2014PhRvD..90b3018L} which contains the bubble evolution along the frequency axis through constructing the light-cone. This we leave for future works as its beyond the scope of the current paper.

While our box size, 150 Mpc, might be small to capture the large scale fluctuations and cosmic variance~\citep{2014MNRAS.439..725I}, we have previously found that our simulation produces a convergent 21cm power spectrum with respect to the volume~\citep[see Figure 8 in][]{2016MNRAS.457.1550H}, such that the 150 Mpc volume produces similar power to that from 300 Mpc volume. This indicates the ionized bubbles distribution in 150 Mpc volumes is similar, on average, to those in large volumes. In addition, it has also been found that such a volume produces a convergent reionization history~\citep{2014MNRAS.439..725I}. However, the resolution (number of cells) is more important for the neural network performance, since higher resolution maps contain more information and structures. Our maps are composed of 200x200 pixels which are able to resolve the small and large scale fluctuations reasonably well.  We leave investigating the network performance in terms of box size and resolution for future works.

\begin{table}
 \caption{The hyperparameters and optimisers used to train the algorithms.}
 \label{setups}

 \begin{tabular}{lllll}
  \hline
   & optimiser & learning rate & batch & cost function\\
  \hline
  \hline
  \textbf{I}  & Nesterov & 0.005 & 128 & $\ell_{1}$ norm \\[2pt]
 \textbf{II}  & Adam & 0.01 & 128 & \verb|rmse| \\[2pt]
 \hline
 \end{tabular}

 \label{optimisers}
\end{table}

\section{Results}\label{results}

To assess how well the algorithms perform in terms of predicting the parameters from learning the input features, we use the coefficient of determination, also known as $R^{2}$ score, which is given by
\begin{equation}\label{r2score}
    R^{2} = 1 - \frac{\sum_{i = 1}^{n}(y_{i}-\hat{y}_{i})^{2}}{\sum_{i = 1}^{n}(y_{i} - \bar{y})^{2}},
\end{equation}
where $\hat{y}_{i}$, $y_{i}$ and $\bar{y}$ are the predicted value, the actual value and the average of all the actual values in the test sample respectively. The numerator of the second term in Equation~\ref{r2score} -- residual sum of squares -- quantifies the variation of the predicted values $\hat{y}_{i}$ around the actual values $y_{i}$, and the denominator accounts for the variation of actual values $y_{i}$ around their mean $\bar{y}$. This metric quantifies the strength of the correlation between the inferred and true values of the parameters, in other words unity $R^{2}$ indicates that the network predictions are identical to the ground truth. The $R^{2}$ also quantifies the fraction by which the predicted variance is less than the true variance. 
For each architecture, we carry out two types of training depending on the input features that the regressors are meant to learn from in order to infer our astrophysical and cosmological parameters (${f_{\rm esc}}, {C_{ \rm ion}}, {D_{ \rm ion}}, {\Omega_{\rm m}}, h, \sigma_{8}$):
\begin{figure*}
\includegraphics[scale=0.75]{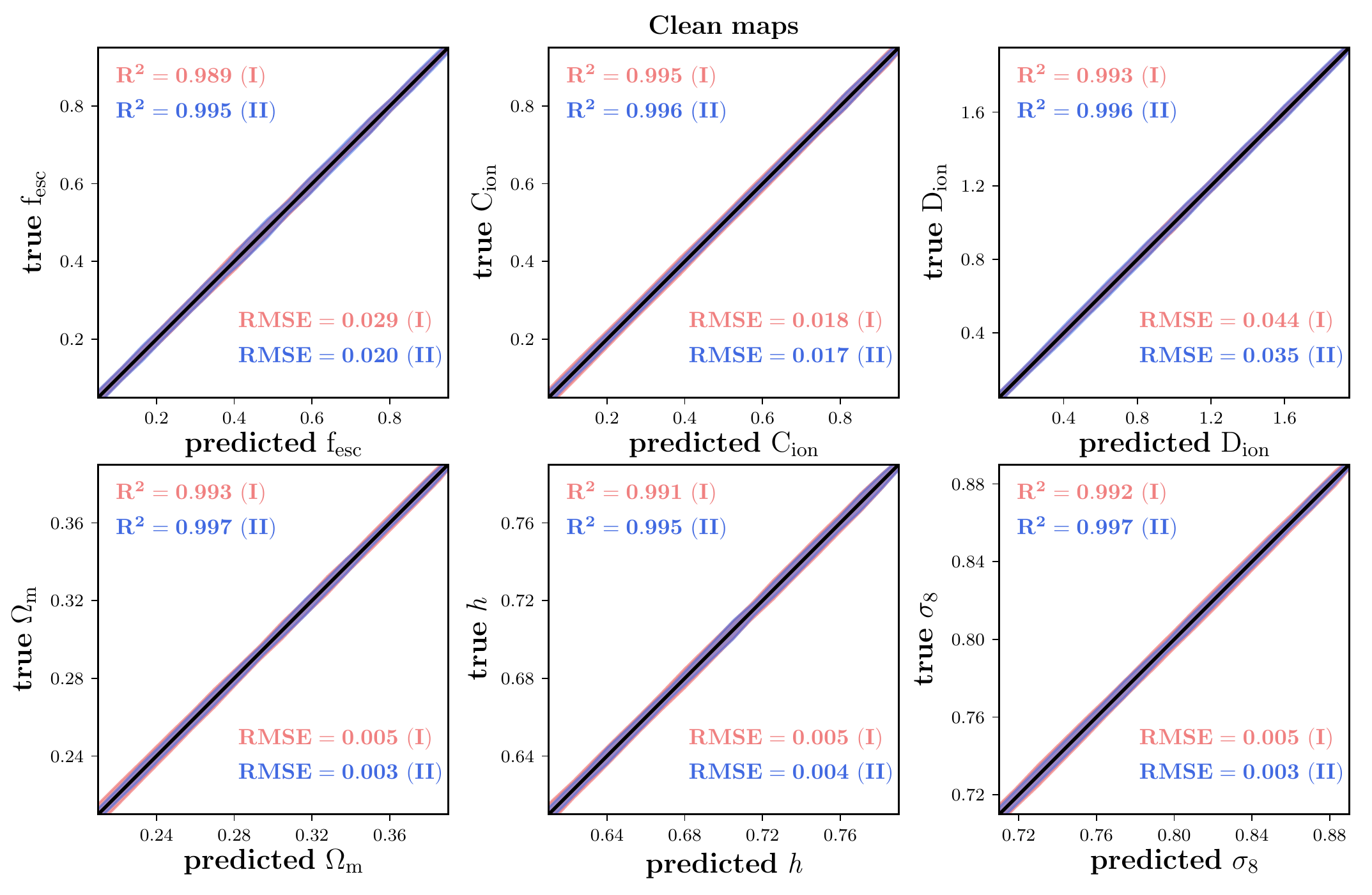}
\includegraphics[scale=0.75]{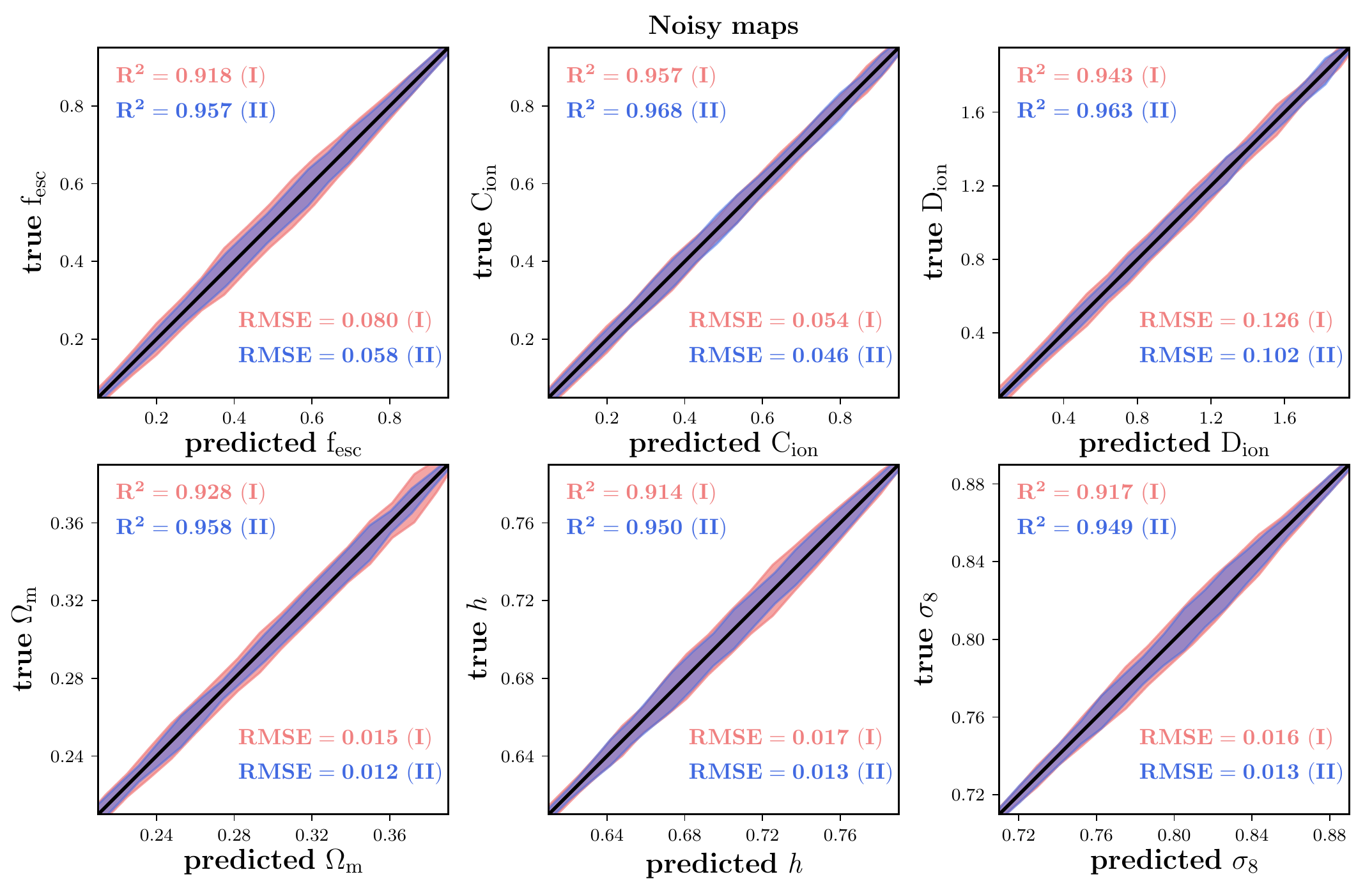}
\caption{Correlation between the true and predicted parameters. On the top two rows, the networks have been trained with the maps without noise, whereas on the bottom two rows, simulated SKA like noise has been injected into the maps which have been used to train the networks. Red and blue shaded areas encompass the
15.9\% and 84.1\%  percentiles (i.e. $\sim$ 1-$\sigma$ level) of the true values given the predictions from \textbf{network I} and \textbf{network II} respectively. Solid black lines represent the identity line \textit{i.e.} true parameters vs true parameters. In all cases, the astrophysical parameters recovery is better than those of the cosmology. Adding the noise reduces the accuracy but the parameter recovery is still promising. The \textbf{network II} outperforms \textbf{network I}, particularly with the mock images, since more complex architecture seems to be needed to extract more information. Large fluctuations are due to low  number statistics.  Coloured version is available online.}
\label{fig_correlation}
\end{figure*}

\begin{itemize}
    \item feature extraction from a simulated 2D 21cm map (\textit{clean/noiseless map} hereafter), this involves training and testing using \textit{clean maps} 
    \item feature extraction from a simulated 2D 21cm~map which was convolved with a simulated SKA like noise (\textit{noisy/mock map} hereafter, see \S\ref{simulation-data}), this consists of training and testing the networks using \textit{noisy maps}.
\end{itemize}

It is worthwhile to mention that we train our networks to predict standardized parameters, meaning that we first subtract the mean and divide by the standard deviation for each array of the parameters. After training, we scale back the predictions to the prior range. Standardizing parameters is important, particularly when the parameters range is different, to prevent the highest parameter range from dominating the loss function. This step is  commonly used in multi-parameters regression deep learning tasks.

\subsection{Learning from \textit{clean maps}}
The top two rows in Figure~\ref{fig_correlation} show the test results when training the networks with the \textit{clean maps}. The red and blue areas   encompass the
15.9\% and 84.1\%  percentiles (i.e. $\sim$ 1-$\sigma$ level) of the true values given the predictions at each bin for \textbf{network I} and \textbf{network II} respectively. Overall, the constraints on each parameter are  very  tight. The high value of the $R^{2}$ score ( $\geq 99\%$ for both \textbf{network I} and  \textbf{network II})  corresponding to each parameter fitting denotes very strong correlation between the true and inferred parameter, suggesting that the algorithms are able to learn the salient features from the data. 
On comparing the performance of the two architectures, their $R^{2}$ score for each fitting suggests that they are in a fairly good agreement, and hence perform equally well. 

\subsection{Learning from \textit{noisy maps}}
The test results after training the algorithms on the \textit{noisy maps} are presented in the bottom two rows in Figure~\ref{fig_correlation}. The constraints on all parameters are slightly weaker as compared to those obtained from training the fitters using the \textit{clean maps}
. The overall decrease in performance denoted by the lower values of $R^{2}$ score corroborates that finding. This can be accounted for by the fact that the relevant features are in this case convolved with noise, therefore extracting them is a bit more challenging. 


Despite being convolved with noise, which essentially causes the quality of their features to degrade, all parameters are successfully recovered with an accuracy of $R^{2} \geq 92\%$ for {\bf network I} and $R^{2} \geq 95\%$ for {\bf network II}, which is remarkably promising.
In contrast to training the algorithms with the \textit{clean maps}, it can be noticed that, overall, \textbf{network II} outperforms \textbf{network I}, as demonstrated by the $R^{2}$ scores of the former, which are a bit higher than those of the latter on all parameters.

\subsection{Dependence on redshift and neutral fraction}

\begin{figure*}
 \centering
    \includegraphics[width=\textwidth]{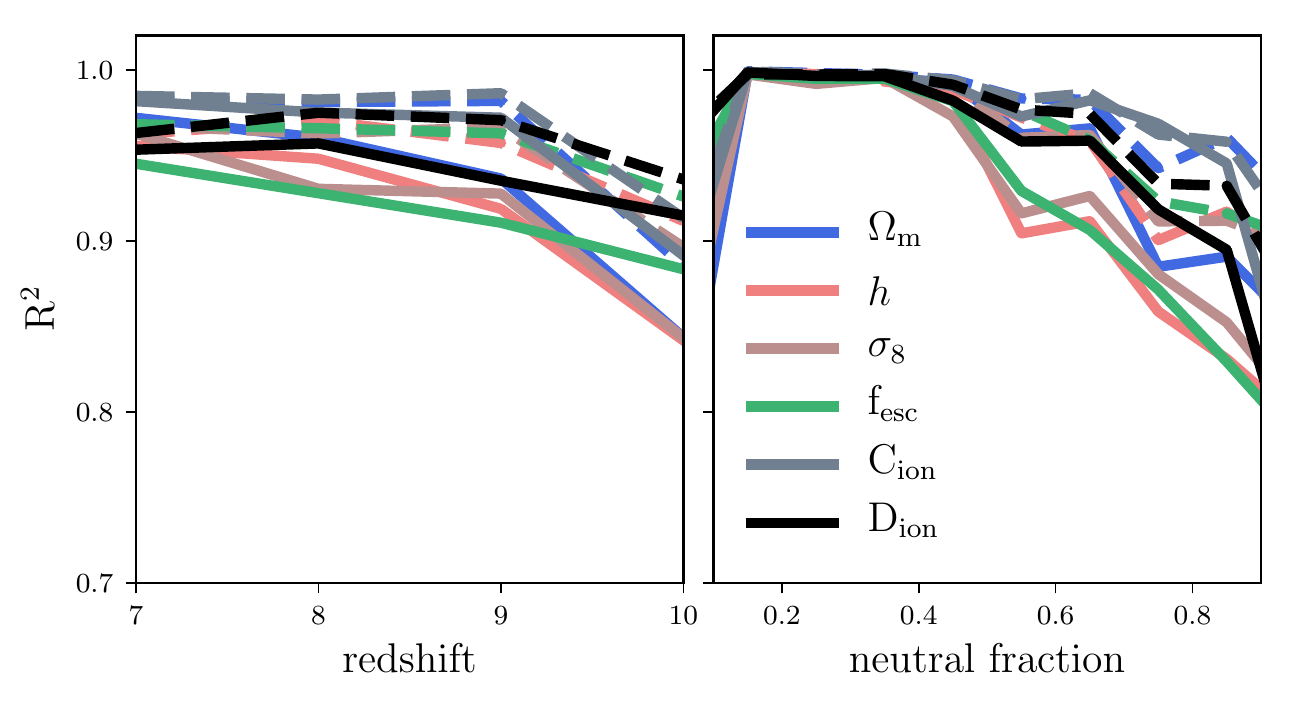}
    \caption{Variation of the resulting coefficient of determination $R^{2}$ as a function of redshift (\textit{left panel}) and neutral fraction (\textit{right panel}). Solid and dashed lines correspond to \textbf{network I} and \textbf{network II} respectively. The accuracy of parameter recovery increases slowly towards low redshift, where the noise is smaller, and rapidly towards low neutral fraction values, where the images features can still be recognized (see Figure~\ref{fig:samples}).  Coloured version is available online. }
    \label{fig_neutral_redshift}
\end{figure*}

\begin{figure*}
\includegraphics[scale=0.75]{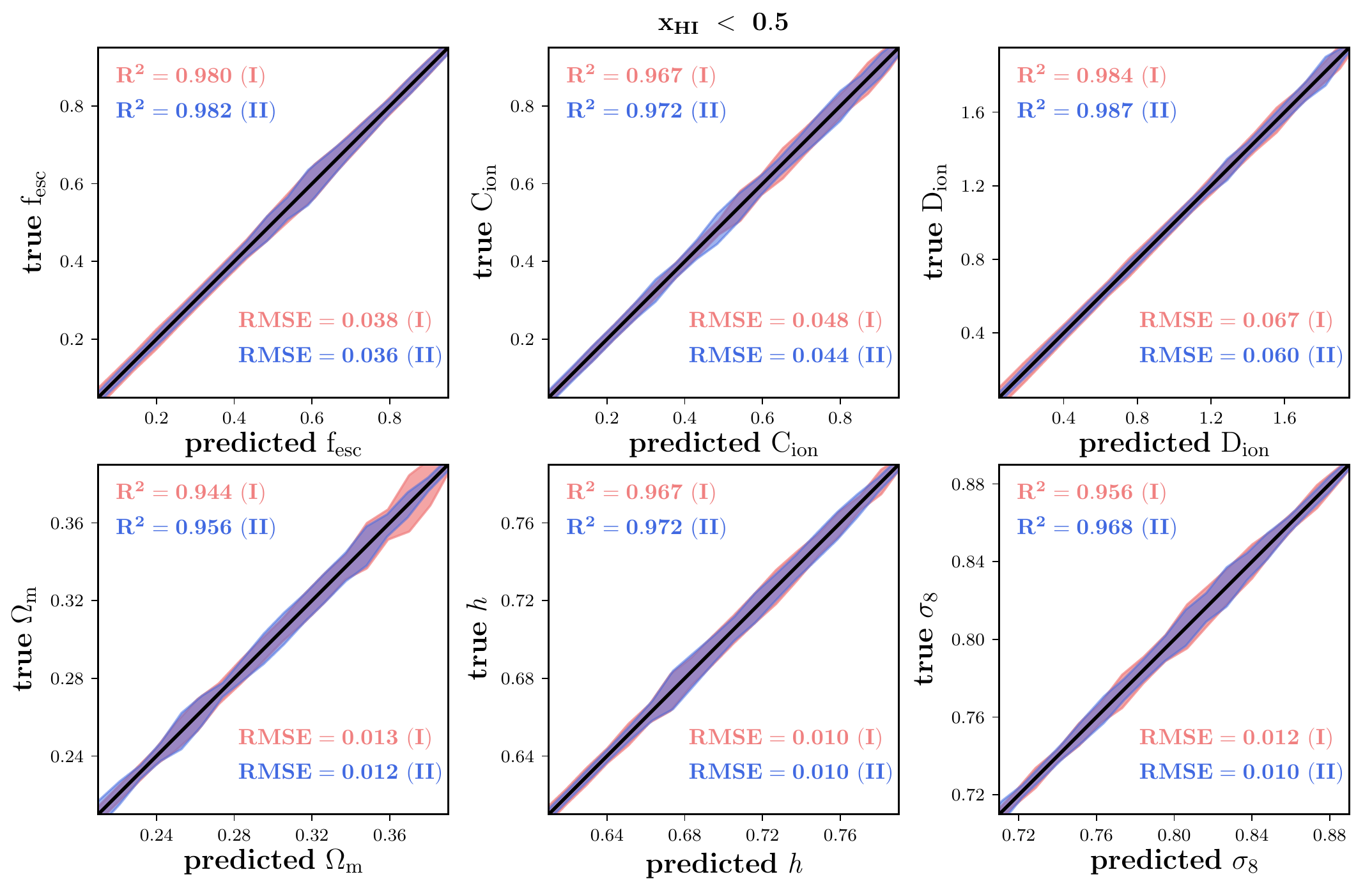}
\includegraphics[scale=0.75]{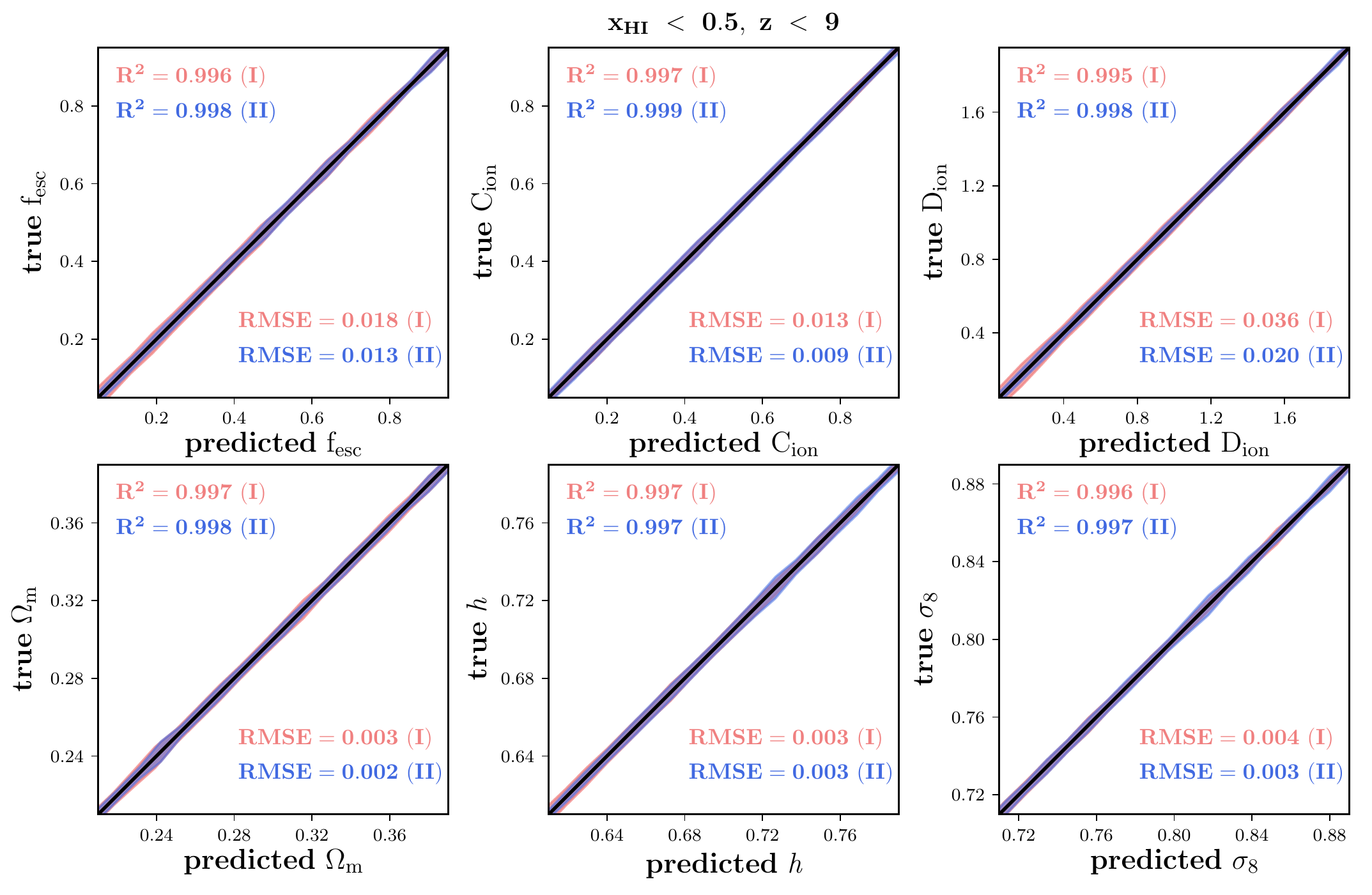}
\caption{Correlation between the actual and the predicted parameters  using the validation sample. On the top two rows, the networks have been trained with the \textit{noisy maps} but a test sample with a neutral fraction $< 0.5$ has been used for predictions. On the bottom two rows, the same \textit{noisy} data have been used to train the networks but some cut on both the neutral fraction $< 0.5$ and redshift $z < 9$ have been applied on the test sample. Red and blue shaded areas encompass the
15.9\% and 84.1\%  percentiles (i.e. $\sim$ 1-$\sigma$ level) of the true values given the predictions from \textbf{network I} and \textbf{network II} respectively. Solid black line represents the identity line \textit{i.e.} true parameters vs true parameters. Imposing constraints on the neutral fraction and redshift of the testing sample increases the accuracy and performance is comparable to the case without including any instrumental effects as seen in top rows in Figure~\ref{fig_correlation}.  Large fluctuations are due to low  number statistics. Coloured version is available online.}
\label{fig_correlation_noisy}
\end{figure*}
In real observations, both foreground contamination and the thermal noise become stronger with increasing redshift. One would then expect some form of dependence of the constraints on redshift. To investigate that possibility, we bin the maps according to their redshift in the test sample and do the predictions by considering each bin separately using the regressors trained with the \textit{noisy maps}. The results presented in the left panel of Figure~\ref{fig_neutral_redshift} suggest the parameter recovery improves with decreasing redshift. 
While \textbf{network II} tends to have a slightly higher accuracy for each parameter as a function of redshift than \textbf{network I}, the dependence on redshift is fairly mild. This weak dependence is due to the fact that there are all possible neutral fraction values at each redshift, without imposing any prior knowledge to the training dataset by allowing certain neutral fraction values for each redshift, following the current reionization history constraints.

As mentioned and seen earlier, the observed features in a 21cm~map are more dependent on $x_{\rm HI}$. To address this effect on the performance of the algorithms, we now bin the slices according their value of $x_{\rm HI}$. It is noticeable in Figure~\ref{fig_neutral_redshift} right panel that the performance of each fitter on all parameters declines with increasing value of the neutral fraction. This is expected, as previously seen in Figure~\ref{fig:samples}, the noise always dominates the ionized regions. When the Universe is highly ionized, the prominent features, which are probably the recombining clumps of the remaining dense gas, can still be seen in the presence of noise. This is in contrast to the case where the Universe is highly neutral, and the bubbles are small. Here, the ionized bubbles extend to much smaller scales where the noise dominates, and hence recognition of the bubbles becomes challenging. At this limit, different realizations (with different sets of parameters) of a highly neutral Universe would look similar. This also explains the rapid increase of the accuracy of parameter recovery towards low neutral fraction values. Similar trends have been recently found with using deep learning to constrain the reionization history~\citep[e.g.][]{mangena19}. It is worthwhile mentioning that this interesting dependence on the neutral fraction cannot be used in future observations since the exact neutral fraction is not known prior observations, although some constraints can be obtained independently from Ly$\alpha$ forest observations~\citep[e.g.][]{2006ARA&A..44..415F}. However, it is beyond the scope of current networks to use this dependence to constrain parameters. It is rather an interesting theoretical finding and consistent with redshift dependence trends since the Universe is highly ionized at low redshifts.

Having established that the constraints are tighter at lower redshift and lower neutral fraction \textit{i.e.} ionised case, we now apply some conditions on the test sample as follows
\begin{itemize}
    \item select examples with $x_{\rm HI} < 0.5$,
    \item select examples with lower neutral fraction at lower redshift, $x_{\rm HI} < 0.5$  and $z < 9$.
\end{itemize}

We show in the top two rows of Figure~\ref{fig_correlation_noisy} the resulting constraints on all parameters when considering maps with $x_{\rm HI} < 0.5$. The results show how the constraints greatly improve by selecting examples with lower neutral fraction from the test sample. For this specific case, the coefficient of determination value $\geq 0.94$ for \textbf{network I} and $\geq 0.96$ for \textbf{network II} on all parameters indicates that the performance of the algorithms, despite considering \textit{mock maps} for their training, is comparable to their performance when trained with \textit{noiseless maps}. 

Restricting ourselves to lower $z$, together with only selecting maps with low neutral fraction, in the test dataset further improves the predictions on all parameters as indicated by $R^{2} \geq 0.99$ for both \textbf{network I} and \textbf{network II}  (see Figure~\ref{fig_correlation_noisy} bottom two rows). Inferring parameters from \textit{noisy maps} at higher redshift is more challenging, since the noise is stronger (see Figure~\ref{fig_neutral_redshift} left panel). Therefore one would expect further improvement of the predictions by combining the two criteria $x_{\rm HI} < 0.5$ and $z < 0.9$.  This is an exciting result for future 21cm surveys that tighter constraints can be obtained from low redshift observations ($z \sim$ 6, 7), where the Universe is highly Ionized. This finding is supported by the fact that the noise is higher at high redshifts, and further confirmed by our additional tests in Appendix~\ref{sec:app2}. Using this technique, the SKA will be able to place their first constraints on the astrophysical and cosmological parameters in the near future and from the first cycle of imaging.

\subsection{Generalisation error}
For the sake of completeness and in order to able to compare our results to other similar studies, we compute, for each parameter, the resulting Root Mean Square Error, {\sc rmse}, as follows:
\begin{equation}
 {\rm RMSE} = \sqrt{  \frac{1}{N} \sum(y_{\rm predicted}-y_{\rm true})^{2}   }\, ,
\end{equation}
where the summation runs through the whole test dataset. This metric, among others, tells us about the generalisation error inherent to our parameter estimation, in other words the level of accuracy\footnote{Not to be confused with accuracy, the metric used in classification tasks.} the fitters can achieve on average when estimating the parameters from encoding the inputs. We show the {\sc rmse} values obtained for each parameter when considering both \textit{noiseless} and \textit{noisy maps}  in Figure~\ref{fig_correlation} (two top rows and two bottom rows respectively). By comparing the {\sc rmse} values resulting from training on \textit{clean maps} and those obtained from training on \textit{mock maps}, we find that in the idealised scenario the prediction is subject to smaller average error for each parameter in contrast to the realistic one. This finding is expected and consistent with our results based on the $R^{2}$ metric in that the inference is more challenging for each parameter when the data considered for training/testing are contaminated by noise. Although the results based on the two metrics are found to be consistent, it is tempting to expect that for any two different parameters, irrespective of the case (\textit{noisy} or \textit{clean maps}), if the $R^{2}$ score of one of them is higher than that of the other, it implies that its {\sc rmse} must be lower. This trend is seen for all parameters as quoted in the legends.

\cite{gupta2018non}, by training a convolutional neural network with $\sim 26$ millions parameters to predict cosmological parameters from simulated noiseless convergence maps, arrived at a generalisation error of $35\times 10^{-3}$ for $\Omega_{{\rm m}}$ and $40.3\times 10^{-3}$ for $\sigma_{8}$. \cite{ribli2018learning} improved the constraints with a different neural network architecture of about $\sim 1.4$ millions parameters, by also using simulated lensing maps, achieving {\sc rmse} = $5.5\times 10^{-3}, 13.5\times 10^{-3}$ for $\Omega_{{\rm m}}$ and  $\sigma_{8}$ respectively. In terms of encoding features from a 2D map using convolutional neural network to infer cosmological parameters, our results -- {\sc rmse} = $5\times 10^{-3}$ and $3\times 10^{-3}$ on $\Omega_{{\rm m}}$ and $\sigma_{8}$ from \textbf{network I} and \textbf{network II} respectively -- are comparable to those obtained in these previous works. More importantly, our results corresponding to the realistic case, with/without imposing constraints (see Figure~\ref{fig_correlation} bottom row and Figure~\ref{fig_correlation_noisy}), where the input maps are \textit{noisy} are very promising and exciting for future 21cm surveys.

\section{Conclusions}\label{conclusion}
We have demonstrated in this work the feasibility of simultaneously inferring both astrophysical and cosmological parameters (${f_{\rm esc}, {C_{\rm ion}}, {D_{\rm ion}}, {\Omega_{\rm m}}, h, \sigma_{8}}$) using 21cm maps from the EoR, considering future \HI~surveys with the SKA. To this end, we have generated thousands of realizations each with a different set of parameters using {\sc simfast21}, then compiled a dataset composed of 2D maps (see \S\ref{sec:train_set} for details). The approach is to train our two proposed algorithms -- convolutional neural network-based -- to extract the features from the maps in order to predict the underlying astrophysical and cosmological parameters. We have considered an optimistic case where we train the networks with \textit{noiseless} simulated maps and a real world-mimicking scenario in which the networks are trained with simulated maps contaminated by simulated SKA-like noise. We have used $R^{2}$ -- coefficient of determination -- as a performance metric.

We summarize our findings as follows:
\begin{itemize}
\item The overall results for the idealised case, with $R^{2}$  $\geq 99\%$ for both \textbf{network I} and \textbf{network II} on all parameters, suggest that the algorithms considered in this work are capable of learning the salient features from the maps in order to constrain the corresponding parameters with a remarkably excellent accuracy. 

\item In a more realistic setup, where maps from observations are subject to noise contamination, the constraints on all parameters are slightly weaker, with an accuracy of $R^{2} \geq 92\%$ for {\bf network I} and $\geq 95\%$ for {\bf network II}. This is expected since disentangling the relevant information from noise is more challenging. It has been found that \textbf{network II}, leveraging the combination of residual layers at lower level and inception module at higher level of the architecture, outperforms \textbf{network I} despite the former's lower capacity. This then points towards deploying similar architectures to \textbf{network II} in a real world scenario.

\item In the case of learning from the \textit{noisy maps}, the predictions are dependent on both the underlying neutral fraction of the map and its distance from an observer. The performance of the methods improves with decreasing neutral fraction and, as foreground contamination is more important at higher redshift, the constraints are tighter at lower $z$. The results obtained from the test sample is $R^{2} \geq 94\%$ with \textbf{network I} and $R^{2} \geq 96\%$ with \textbf{network II} when only selecting maps with $x_{\rm HI} < 0.5$. Recovery improves with imposing constraints on both neutral fraction and redshift ($x_{\rm HI} < 0.5$ at lower $z$ $< 9$), resulting in $R^{2} > 99\%$ with both \textbf{network I} and \textbf{network II}. This indicates that even in the presence of \textit{noise} in maps, our methods can still estimate the relevant parameters to an excellent level of precision, which is indeed quite promising.

\item We have computed the prediction error on average -- {\sc rmse} -- on each parameter in both optimistic and realistic cases. It has been found that the {\sc rmse} is smaller in the former, in good agreement with the results when using the coefficient of determination as a performance metric. Compared to other previous works, our approach has also shown a great potential for inferring the underlying parameters of what is observed in future cosmological experiments, such as \HI~intensity mapping.
\end{itemize}

We here considered a redshift range that is consistent with an early reionization scenario, which has been increasingly favoured by galaxy-dominated models of reionization, although more recent work by~\citet{2019MNRAS.485L..24K} shows that galaxies can produce low optical depth and a late reionization scenario. However, late reionization as usually favoured by AGN-dominated scenarios is currently disfavoured~\citep[e.g. see][]{2017MNRAS.472.2009Q,2018MNRAS.473..227H,2018MNRAS.473.1416M,2018MNRAS.474.2904P}. Regardless of the redshift range, the main result, that the accuracy increases with decreasing redshift and neutral fraction, would qualitatively remain valid if lower redshifts are included in this study, such as z=$5,6$, since the instrumental effects are always higher at a higher redshift. 
 
In our analysis, we have generated our training samples based on 1,000 different reionization simulations to constrain 6 parameters. For instance,~\citet{gupta2018non},~\citet{2019ApJ...880..110L}, and~\citet{2019MNRAS.484..282G} have used 96, 1,000, 10,000 model evaluations to constrain 2, 1, 3 parameters.~\citet{2018MNRAS.475.1213S} further have shown that 100 model evaluations is sufficient to constrain 3 parameters. In comparison to these works, the number of simulations used in this study is low, which limits the presented results. The prior range assumed in this study is also small (i.e. 0.2 - 2) which places additional limitation to our results. Higher accuracy than reported in this study is expected with larger training samples and more model evaluations, which we will explore in future works.
 
Our results are entirely limited to the set of assumptions and approximation implemented in our 21cm instrument simulation.
A more refined and sophisticated recipe to account for all of the implemented instrumental effects, such as the angular resolution, foreground cleaning and thermal noise, might alter our concluding remarks. The approximation and assumptions implemented in the semi-numerical simulations, through the use of the excursion set formalism to identify the ionized regions, as well as the choice of our dynamic range and resolution, place additional limitations to the presented results. While limited to the SKA, our analysis can be easily extended to include instrumental effects from other 21cm surveys such as HERA and LOFAR, which we leave for future works to perform a detailed comparison between different array designs and different observing strategies. Inferring parameters from the 3D light cones might improve recovery in the presence of noise without the need to impose constraints on the neutral fraction or redshift. Our analysis also can be easily extended to include all of the astrophysical parameters from the source and sink models, and all cosmological parameters, which we leave for future works.  

This study has not only highlighted the constraining power of our methods, probing deep into EoR in the near future with the arrival of more advanced \HI~instruments like SKA, but also shown how future 21cm surveys and \HI~intensity mapping can help break the degeneracy between models by combining them with other experiments, such as {\it Planck}, to better the constraints on cosmological parameters in an era of precision cosmology.

\section*{Acknowledgements}
The authors acknowledge helpful discussions with Laura Boucheron, Kristian Finlator,  Jon Holtzman, Tumelo Mangena, James McAteer,  Mario Santos, and Rene Walterbos.  We particularly thank the anonymous referee for their constructive comments which have improved the paper quality significantly. Simulations and analysis were performed at UWC's {\sc Pumbaa}, IDIA/{\sc Ilifu} cloud computing facilities, and NMSU's {\sc DISCOVERY} supercomputers. This work also used the Extreme Science and Engineering Discovery Environment (XSEDE), which is supported by National Science Foundation grant number ACI-1548562, and computational resources (Bridges) provided through the allocation AST190003P. SA acknowledges financial support from the {\it South African Radio Astronomy Observatory} (SARAO) and is grateful to Sean February and Martin Slabber from the \textit{Science Data Processing} (SDP) team at SARAO for their strong technical support with regards to the computing resources. CCD thanks the LSSTC Data Science Fellowship Program, which is funded by LSSTC, NSF Cybertraining Grant \#1829740, the Brinson Foundation, and the Moore Foundation; Their participation in the program has benefited this work. CCD acknowledges funding from the New Mexico Space Grant Consortium Grant \#NNX15AL51H.





\bibliographystyle{mnras}
\bibliography{cnn21cm_bib.bib}


\newpage

\appendix

\section{The loss function evolution during training}\label{sec:app1}
\begin{figure*}
    \includegraphics{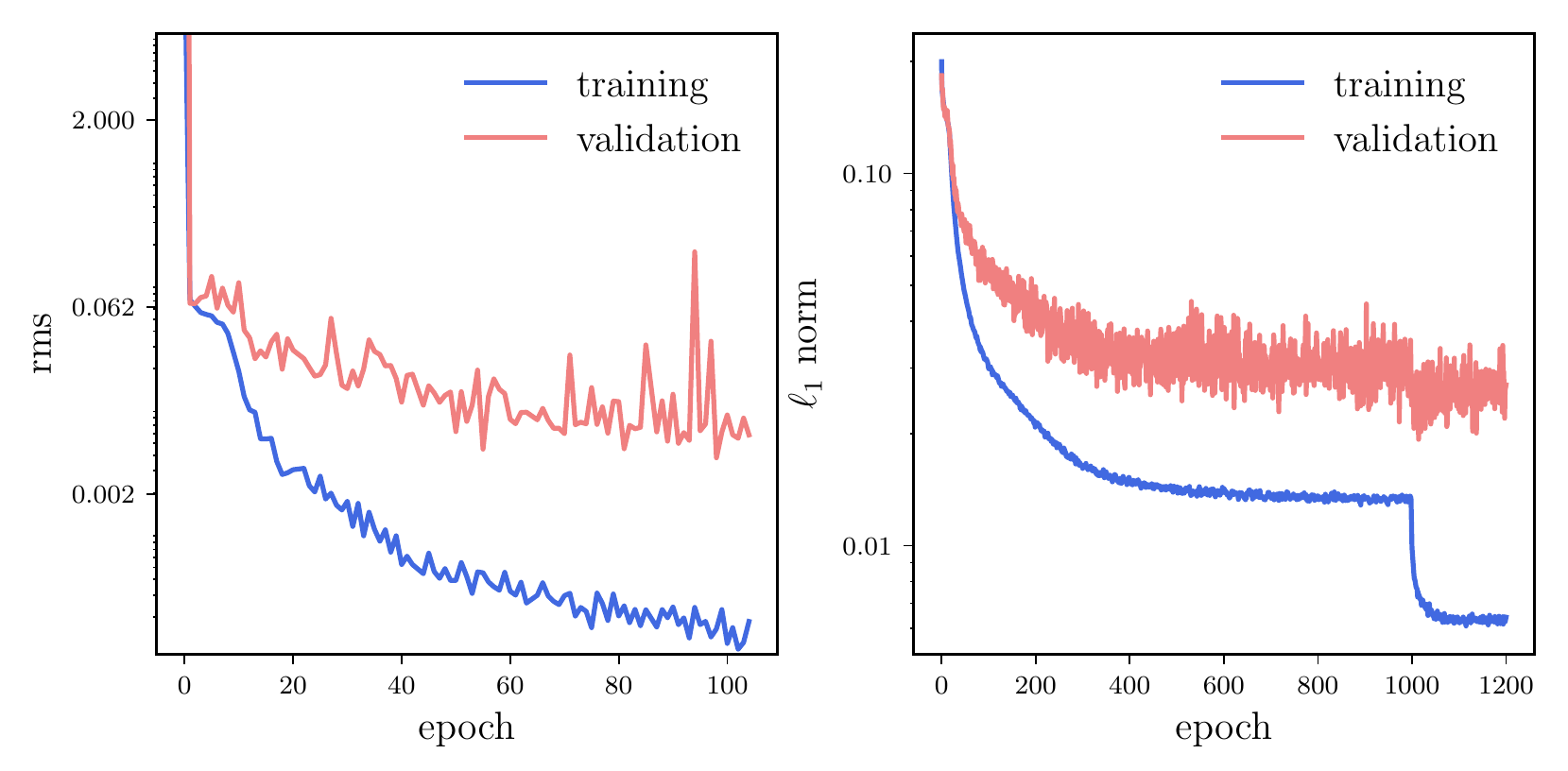}
    \caption{ \textit{Left panel}: Progression of the training of {\bf network I} where the loss function {\sc rms} varies as a function of training epoch. \textit{Right panel}: Progression of the training of {\bf network II} showing the loss function $\ell_{1}$ norm as a function of number of epochs. Coloured version is available online. The two plots are related to the training on the noisy data.}
    \label{fig:progress}
\end{figure*}

Figure~\ref{fig:progress} shows the loss evolution of {\bf  network I} (left) and {\bf II} (right) for training (blue) and validation (red) samples as a function of training epoch for the case of training on the noisy data set.  In both cases, the loss is decreasing as training progresses, which indicates a reduction in the error rate and predictions are approaching the target labels. The fluctuations in the training and validation curves are due to random selection of batches during training. Regardless of these fluctuations, the loss evolution for validation converges and stays constant on average, which indicates that the networks are not over fitting. It is worth noting that the sudden drop in training/validation error while training {\bf network II} is owing to the fact that the learning rate is updated to 10\% of its initial value in order to escape the plateau. 

\section{Redshift evolution}\label{sec:app2}
Our main result which suggests that the accuracy increases with decreasing redshift has been derived from a model trained on mixed maps from all redshifts. To confirm whether accuracy increases towards low redshift, we here perform additional learnings by restricting the training sample to have maps only from the minimum (z=7) or maximum (z=10) redshifts considered in this study (referred to as only). We also compare with predictions at these redshifts from training with the whole dataset, including all other redshifts (referred to as whole), for the case of noisy maps as reported in Table~\ref{tab:accuracy_redshift}. In all cases, we find that the accuracy at z=7 is always higher than that of at z=10. This shows, regardless of training with whole mixed maps or maps at a given redshift, the qualitative trend that the accuracy increases towards low redshifts is still seen as summarized in Table~\ref{tab:accuracy_redshift}. In addition, the quantitative results are also similar with a minimal difference of about $\lesssim$ 2\% of accuracy for some parameters as summarised in Table~\ref{tab:accuracy_redshift} for training with maps at individual redshifts (only) versus those derived using a trained model on mixed maps (whole). Such a minimal difference is expected due to the different number of samples used in the case of ``only'' versus ``whole'' tests. This shows that our networks are successful to recover the same qualitative and quantitative results without explicitly including the redshift information as an input to the network (e.g. fitting parameters to four maps from the four different redshifts, $z=10-7$).

\begin{table*}
 \caption{Networks accuracy comparison between training only with dataset from z=7 and z=10 (referred to as only) versus predicting at these redshifts from training with whole dataset (including all other redshifts, referred to as whole), for the case of noisy maps. For all parameters with all networks, accuracy increases towards low redshift.}
 \label{tab:accuracy_redshift}
 \begin{adjustbox}{width=6in,center}
\begin{tabular}{c c c c c c c c c }
  & \multicolumn{4}{|c|}{\bf Network I} & \multicolumn{4}{|c|}{\bf Network II} \\[2pt]
   \hline
   &  $z=10$ (only)  & $z=10$ (whole)  & $z=7$ (only) &  $z=7$ (whole)   & $z=10$ (only)  &   $z=10$ (whole)   & $z=7$ (only) & $z=7$ (whole)\\[2pt]
   \hline
  $\Omega_{m}$  & 0.86 & 0.84 &0.97& 0.97 & 0.86& 0.88&  0.96 & 0.98\\[2pt]
  $h$           & 0.87 & 0.84 &0.95& 0.95  & 0.88& 0.91& 0.95 & 0.96\\[2pt]
  $\sigma_{8}$  & 0.86 & 0.84 &0.95& 0.96 & 0.88& 0.89& 0.96 & 0.96\\[2pt]
  $f_{\rm esc}$ & 0.90 & 0.88 &0.95& 0.94 & 0.90& 0.92& 0.95 & 0.96\\[2pt]
  $C_{\rm ion}$ & 0.91 & 0.89 &0.98& 0.98 & 0.90& 0.91& 0.98 & 0.98\\[2pt]
  $D_{\rm ion}$ & 0.93 & 0.91 &0.95& 0.95 & 0.92& 0.93& 0.95 & 0.96\\[2pt]
 \hline
 \end{tabular}
 \end{adjustbox}
 \label{optimisers}
\end{table*}

\end{document}